\documentclass[reprint,amssymb, amsmath, aip,cha,onecolumn]{revtex4-1}
\usepackage{graphicx}
\usepackage{color}
\usepackage{epsfig}
\usepackage{bm}%
\usepackage[colorlinks=true,linkcolor=blue]{hyperref}%
\expandafter\ifx\csname package@font\endcsname\relax\else
 \expandafter\expandafter
 \expandafter\usepackage
 \expandafter\expandafter
 \expandafter{\csname package@font\endcsname}%
\fi
\begin{document}
\title{Experimental investigation of dynamical structures formed due to a complex plasma flowing past an obstacle}%

\author{S. Jaiswal}
%\author{S. Jaiswal}%
\thanks{Presently at Physics Department, Auburn University, Auburn, AL 36849 USA}
\email{surabhijaiswal@gmail.com}
\affiliation{Institut f\"ur Materialphysik im Weltraum, Deutsches Zentrum f{\"u}r Luft- und Raumfahrt (DLR), 82234 We{\ss}ling, Germany.}%
\affiliation{Institute For Plasma Research, Bhat, Gandhinagar,Gujarat, India, 382428}%
\author{M. Schwabe}
\affiliation{Institut f\"ur Materialphysik im Weltraum, Deutsches Zentrum f{\"u}r Luft- und Raumfahrt (DLR), 82234 We{\ss}ling, Germany.}%
\author{ A. Sen}
\author{P. Bandyopadhyay}
%\eaddress{pintu31@gmail.com}

%\author{A. Sen}
%\author{P. Bandyopadhyay}
%\affil[aff1]{Deutsches Zentrum f{\"u}r Luft- und Raumfahrt (DLR), 82234 We{\ss}ling, Germany.}
%\affil[aff2]{Institute for Plasma Research, Bhat, Gandhinagar, Gujarat 382428, India.}
%\affiliation{Deutsches Zentrum f{\"u}r Luft- und Raumfahrt (DLR), 82234 We{\ss}ling, Germany.}%
\affiliation{Institute For Plasma Research, Bhat, Gandhinagar,Gujarat, India, 382428}%
\date{\today}
%**************************************************************
%#####################################################################################                  ABSTRACT
%************************************************************************************************
\begin{abstract}
We report experimental observations of the dynamical behavior of a complex plasma flowing past a spherical obstacle. The investigation has been carried out in a $\Pi$-shaped DC glow discharge experimental device where a dust cloud of kaolin particles in a background Argon plasma is made to flow in a controlled fashion by regulating the mass flow of the neutrals. A spherical metal object vertically mounted on the cathode tray acts as an obstacle to the flow. The flowing dust particles are repelled by the electrostatic field of the negatively charged sphere and a microparticle free region (dust void) is formed surrounding the obstacle. The distant dust particles are attracted towards the floating obstacle and reflected back when they get to a certain distance, causing a ring shaped structure around the obstacle. We characterize the shape of this structure over a range of dust flow speeds and obstacle biases. For a supersonic flow of the dust fluid a bow shock is seen to form on the upstream side of the negatively biased sphere, while the downstream side shows the generation of wave structures for a particular range of flow velocities when the Reynolds number $R_e \gtrsim 50$. %The wave structure is reminiscent of the early formation stage of a Von-K\'arm\'an vortex street. 
The flow generated structures can be physically understood in terms of the dust dynamics under the combined influence of the ion-drag force, the neutral streaming and the electric force.
\end{abstract}
%%%%%%%%%%%%%%%%%%%%%%%%%%%%%%%%
\maketitle
%%%%%%%%%%%%%%%%%%%%%%%%%%%%%%%%%%%%%%%%%%%%%%%%%%%%%%%%%%%%%%%%%%%%%%%%%%%%%%%%%%%%%       INTRODUCTION
%%%%%%%%%%%%%%%%%%%%%%%%%%%%%%%%%%%%%%%%%%%%%%%%%%%%%%%%%%
\section{Introduction}\label{sec:intro}
A study of the nature of fluid flow past an obstacle has attracted the attention of many researchers \cite{Batchelor} over the years and continues to be of fundamental interest even to this day. The interest arises both from its significant importance in gaining understanding of various phenomena in biology, oceanography, and atmospheric dynamics, as well as for its variety of  practical implications in astrophysical situations and various other engineering applications \cite{Saffman, Zhang, Sheard, Dynnikova}. As an example, a study of air flow past tall buildings helps in determining means of suppressing vortex shedding in order to reduce the total drag and to minimize unsteady aerodynamic loads. Likewise a knowledge of the flow in tube banks in heat exchangers \cite{ying}
is necessary to increase the heat transfer efficiency and to decrease the induced vibration and fatigue damage. 
In space, the magnetic field of the earth acts as an obstacle for the solar wind, and the interaction between them results in the generation of the well known bow shock in the upstream region of the planet \cite{Ness}. In the \lq\lq Mars Global Surveyor (MGS)'' mission, among others, a similar phenomenon has been observed due to the interaction of the solar wind with Mars \cite{mars}. It was found that the solar wind slowed down to subsonic speeds in the vicinity of Mars at a parabolic surface, the \lq\lq bow shock'' \cite{mars}. In addition to this, the flow of the solar wind becomes chaotic due to fluctuations in the magnetic field, and an instability arises in the downstream side. This fascinating phenomenon of dynamical structure formation in the downstream direction has also been investigated in variety of other media \cite{samsonov, havnes, melzer}.  \par
%It is well documented in previous investigation that 
Several past investigations in fluid mechanics have revealed that the fluid flowing around an obstacle generates a pair of vortices on the downstream side of the obstacle. Depending on the flow velocity, a growth, detachment, and shedding of the vortices from the obstacle have been seen to occur leading to the formation of a vortex street in the shadow of the obstacle \cite{Saffman, Zhang, Sheard, Dynnikova, karman}. \par 
A dusty plasma consisting of electrons, ions, and charged micron or sub-micron-sized particles (``dust'') offers an excellent environment to study phenomena associated with flow past an obstacle \cite{morfill1, morfill2}. This is due to the fact that the charged microparticles, interacting with each other through electric fields mediated by the plasma, can be considered as an essentially one-phase system like fluids. This unique feature of complex plasmas can be employed to study fluid dynamics at the individual particle level by using simple optical diagnostics such as laser illumination and video imaging through high speed ccd cameras along with suitable image analysis techniques \cite{sergey, samsonov2, heidemann, jaiswal1}. \par
 Such a diagnostic convenience has facilitated many experimental investigations of flow phenomena in dusty plasmas 
 including flow past obstacles \cite{morfill1, thompson, ed, samsonov2, Klindworth, nakamura, meyer, jaiswal2, jaiswal3}. Thompson \textit{et al.} \cite{thompson} experimentally investigated the interactions between an electrically floating metallic rod and charged microparticles either by keeping the rod at rest or by moving it through a dusty plasma with velocities of the order of or much faster than the dust acoustic velocity. They did not observe any structure formation for the higher velocity cases. Thomas \textit{et al.} \cite{ed} studied the formation of a dust void and the scaling of its size with the biasing voltage of a probe in a dusty plasma. Samsonov \textit{et al.} \cite{samsonov3} experimentally demonstrated the interaction of a negatively biased wire with a lattice of microparticles. They showed that the attraction of distant particles and the repulsion between near ones depends on the magnitudes of the ion drag and the electrostatic force respectively. Klindworth \textit{et al.} \cite{Klindworth} observed a dust free region around a Langmuir probe in a radio- frequency (RF) produced complex plasma under microgravity conditions. The balancing of the electric field force and the ion drag force was considered to be the dominant mechanism for void formation in their experiment. Morfill \textit{et al.} \cite{morfill1} performed an experiment on dust fluid flow around a lentil - shaped obstacle using a complex plasma in the liquid state. They observed a mixing layer between the flow and the wake that exhibited an instability growing on a scale much smaller than the hydrodynamic scale. Later, in 2011, Heidemann \textit{et al.} \cite{heideman2} carried out a similar experiment and observed a shear instability. They reported that the two-stream interface between flow and stagnation zone breaks down into a well-developed multi-stream network due to Rayleigh-Taylor-type perturbations. Saitou \textit{et al.} \cite{nakamura} observed the formation of a bow shock as a consequence of a supersonic flow of charged microparticles around a negatively biased needle. They performed their experiment in a specially designed dusty plasma chamber where the flow was generated by tilting the device with the help of a mechanical jack. Meyer \textit{et al.} \cite{meyer} performed an experiment on transient bow shock formation along with a teardrop-shaped wake region on the downstream side by a supersonic dust fluid incident on a biased cylinder. Very recently, precursor solitons and shock waves have been observed for supersonic dust fluid flow over a wire horizontally placed over the cathode plate in a dusty plasma experimental device dedicated to flowing plasma experiments \cite{jaiswal2, jaiswal3}. Emergence of secondary bow shocks in the downstream of an obstacle is observed in a numerical study by Charan \textit{et al.} \cite{harish1}.\par 
 Most of these past flow experiments in dusty plasmas have focused their attention on a particular single phenomenon such as a void formation or the emergence of a shock structure. A detailed study of the various dynamical structures as a function of the dust fluid flow velocity and bias voltages of the obstacle is lacking. In this paper we report on such a study carried out in the direct current (DC) glow discharge Dusty Plasma Experimental (DPEx) device \cite{jaiswal1}. An isolated metallic sphere is used as an obstacle to the flow of dust fluid that is generated by changing the gas flow rate in a controlled manner. The obstacle is kept either at a floating potential or biased negatively at different voltages. The experiment is carried out at different Reynolds numbers corresponding to different dust flow speeds. A variety of dynamic structures are seen to emerge starting with a void formation due to the repulsion of the dust particles by the floating sphere. The distant dust particles are attracted and get trapped in the potential of the sphere to form a spherical halo around the obstacle. The shape of this halo can be affected by a change in the flow velocity. At higher flow velocity the distribution of dust become asymmetric with respect to upstream and downstream region of the obstacle. For a negatively biased obstacle, a bow shock like structure is found to form in the upstream side whereas an instability grows and generates a propagating wave structure in the downstream side when the flow was supersonic.
 % For a supersonic flow a bow shock like structure is found to form in the upstream side of the negatively biased obstacle whereas an instability grows and generates a propagating wave structure in the downstream side. 
 These and other dynamical characteristics of the flow generated patterns are physically explained on the basis of the relative strengths of the various competing forces acting on the dust particles, namely, the repulsive force due to the charged sphere, the ion drag force and the effect of the neutral streaming.\par
The paper is organized as follows. In the next section (Sec.~\ref{sec:setup}), we present the experimental set-up in detail. In Sec.~\ref{sec:results}, we discuss the experimental results on the excitation of various dynamical structures. Sec.~\ref{sec:conclusion} provides some discussion and concluding remarks.
 %%%%%%%%%%%%%%%%%%%%%%%%%%%%%%%%%%%%%%%%%%%%%%%%%%%%%%%%%%%%%%%%%%%%%%%%%%%%%%%%%%%%%%%%%%%%%
 %&&&&&&&&&&&&&&&&&&&&&&&&&&&&&&&&&&&&&&&&&&&&&&&&&&&&&&&&&&&&&&&&&&&&&&&&&&&&&&&&&&&&&&&&&&&&&&&&&&&&&&&&&&&&&&&
 %***********************************************************************************************************************
\section{Experimental Setup and method}\label{sec:setup}
\begin{figure}[!ht]
\centering{\includegraphics[scale=0.65]{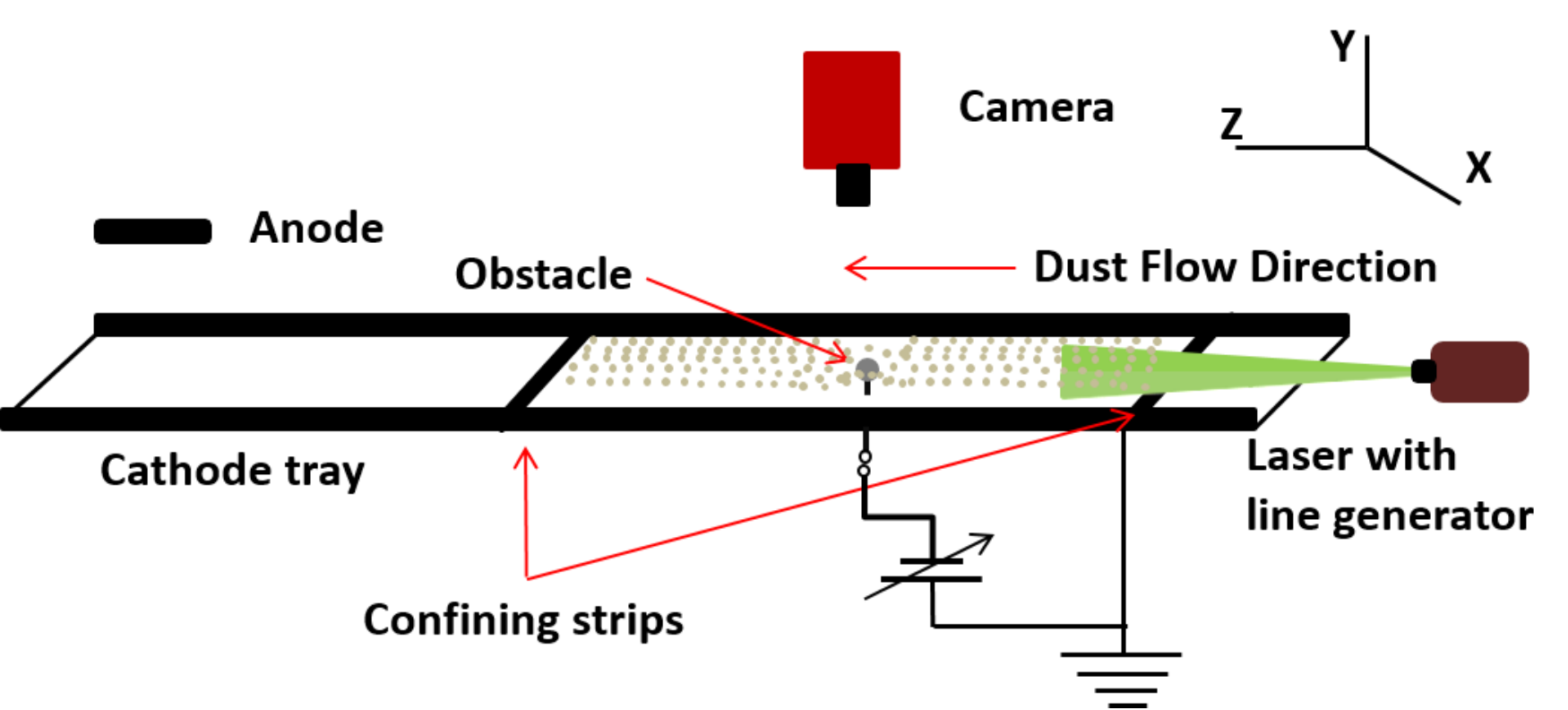}}
\caption{\label{fig:setup} A schematic of the Dusty Plasma Experimental (DPEx) Device.}
\end{figure}

The experiments have been performed in a table top Dusty Plasma Experimental (DPEx) device (shown schematically in Fig.~\ref{fig:setup}). A detailed description of the experimental set-up has been reported elsewhere (see ref.\cite{jaiswal1}). The vacuum chamber consists of a $\Pi$-shaped glass tube with a horizontal section of 8 cm inner diameter and 65 cm length. This section is utilized for the experimental study of the dusty plasma. The two vertical sections of 30 cm length and of same diameter, are utilized for pumping, gas inlets, anode connection, and for other functional access. A stainless steel (SS) disc electrode of 3 cm diameter (insulated with ceramic from the back side) serves as an anode whereas a grounded SS tray of 40 cm $\times$ 6.1 cm$\times$ 0.2 cm, placed on the horizontal area (Z-axis), is used as a cathode. Two SS strips are placed approximately $~25~$cms apart on the cathode to confine the dust particles in the axial direction by means of their sheath electric fields. A silver ball of 5 mm diameter, attached to a ceramic rod for support, is mounted vertically in between these two strips such that the top of the sphere is at a height of $1.8~$cm from the cathode base. The potential of the obstacle is either kept floating or set at different potentials ranging from +250 V to +280 V with respect to the ground, which is negative with respect to the plasma potential, $V_p \approx +345 ~$V. Before closing the experimental chamber, micron sized kaolin powder (with a size dispersion ranging from $2$ to $4$ microns) is sprinkled on the cathode tray.\par
%

%%%%%%%%%%%%%%%%%%%%%%%%
%%%%%%%%%%%%%%%%%%%%%%%%
 The experimental chamber is evacuated to a base pressure of $10^{-3}$ mbar by a rotary pump. Argon gas is then flushed into the chamber through a mass flow controller which is connected in the horizontal section of the glass tube and interfaced with the computer. The flow controller is the heart of the experimental setup as it is utilized for manipulating the dust cloud by changing the gas flow in a well controlled manner.
To remove any kind of impurity, the chamber is flushed several times with Argon gas and pumped down to base pressure. Finally the background pressure is set to a working pressure of $P=0.110-0.120~$mbar by adjusting the pumping speed and the gas flow rate. In this condition, the gate valve (connected to the mouth of the pump) is  opened approximately to 20\% and the mass flow controller is opened by $10\%$ (corresponding to a flow of 27.5 ml$_s$/min) to maintain a constant pressure. A DC glow discharge plasma is then produced in between the anode and the grounded cathode by applying a voltage, $V_a = 400~$V. The applied voltage is then reduced to $350-370~$V at a discharge current of $I_p \sim 4-8~$mA. The plasma parameters such as ion density ($n_i$), electron temperature ($T_e$) and plasma/floating potentials are measured initially using a single Langmuir probe and emissive probes in the absence of dust particles. The typical experimentally measured values of the plasma parameters (at $P = 0.110-0.120~$mbar and $V_a \approx 360$ V) are $n_{i} \approx 10^{15}~$m$^{-3}$, $T_{e} \approx 4-5~$eV. More details about the complete evolution of the plasma parameters over a broad range of discharge conditions can be found in Ref. \cite{jaiswal1}. The ions are assumed to be at room temperature, i.e., $T_{i} \approx 0.03~$eV \cite{thompson}. For the given discharge parameters, a dense dust cloud of kaolin particles is found to levitate near the cathode sheath boundary. The levitated dust particles are illuminated by a green diode laser placed at a height of 1.7 cm from the cathode plate. The laser beam has a width of about 1.5 mm  and the scattered light from the dust cloud is captured by a fast CCD camera (122 fps) with 1000 pixels $\times$ 350 pixels resolution. The height of the laser beam has been chosen to ensure a continuous view of the dust cloud in the course of the entire flow experiment. The vertical extent of the dust cloud is therefore seen to be $\geq 1.5 mm$ and the cloud is located at a height such that it can intercept the obstacle during its flow.
%We confirm this from a image sequence which recorded the journey of the cloud from its stationary position to flowing past through the obstacle . 
From the analysis of the video images the dust density is estimated to be $n_{d} \approx 10^{11}~$m$^{-3}$ \cite{jaiswal1, jaiswal3}. The dust mass is $m_{d} \approx (1-9)\times 10^{-14}~$kg and the average charge on a dust particle (inferred from the plasma parameters and the particle size) is approximately $Q_{d} \approx 10^{4}~$e \cite{barken}. The electron density is obtained from the quasi-neutrality condition by taking into account the dust contribution. The dust temperature is taken as room temperature, $T_d \approx 0.03$~eV. Based on these values, the typical magnitude of the linear phase velocity of dust acoustic waves (DAWs) \cite{merlino_daw} for our experimental conditions turns out to be $v_{ph} \sim  3-5~$cm/sec. 
\par
To run the experiment,  we first adjust the pumping speed and gas flow rate in a precise way to achieve a stationary dust cloud on the right side of the obstacle. 
This equilibrium condition is achieved at $\sim 10\%$ opening of the mass flow controller and $\sim $ 20$\%$ opening of the gate valve. From this equilibrium condition, if we now change the gas flow rate, the particles are found to move either towards the pump (in case of decreasing flow rate) or away from the pump (in case of increasing flow rate). A detailed description of different means of flow generation and measurement techniques has been reported in previous publications by Jaiswal \textit {et al.} \cite{jaiswal4, jaiswal1}. In the particular set of experiments presented here, the flow of the dust fluid is generated by reducing the gas flow rate in steps of 1$\%$ which corresponds to decreases in flow rate of 2.75 ml$_s$/min. In this way the induced flow of neutrals cause the microparticles to stream always from the right to the left of the obstacle. With the change of sudden gas flow, the particles are initially accelerated towards the pump from the equilibrium position. After traveling a distance of less than 1 cm, almost all the particles are found to achieve a constant velocity, which is approximately equal to background neutral velocity. This velocity is found to remain uniform over a substantial region of the device until the time that the cloud runs out of dust particles. It is also worth mentioning that when we induce such a flow in the absence of the potential obstacle we do not observe any flow vortices or other complex gas flow structures in the video images of the dust particles. This indicates that the gas flow is of a laminar nature. %and the observed flow structures in the experiments are solely due to the interaction of the dust flow with the obstacle.} 
In the following, the given flow rates of the background gas correspond to the change with respect to the equilibrium condition (i.e. without dust flow).
%%%%%%%%%%%%%%%%%%%%%%%%%%%%%%%% 
\section{Experimental Observation}\label{sec:results}
%%%%%%%%%%%%%%%%%%%%%%%
\begin{figure*}[!ht]
\centering{\includegraphics[width=0.9\textwidth]{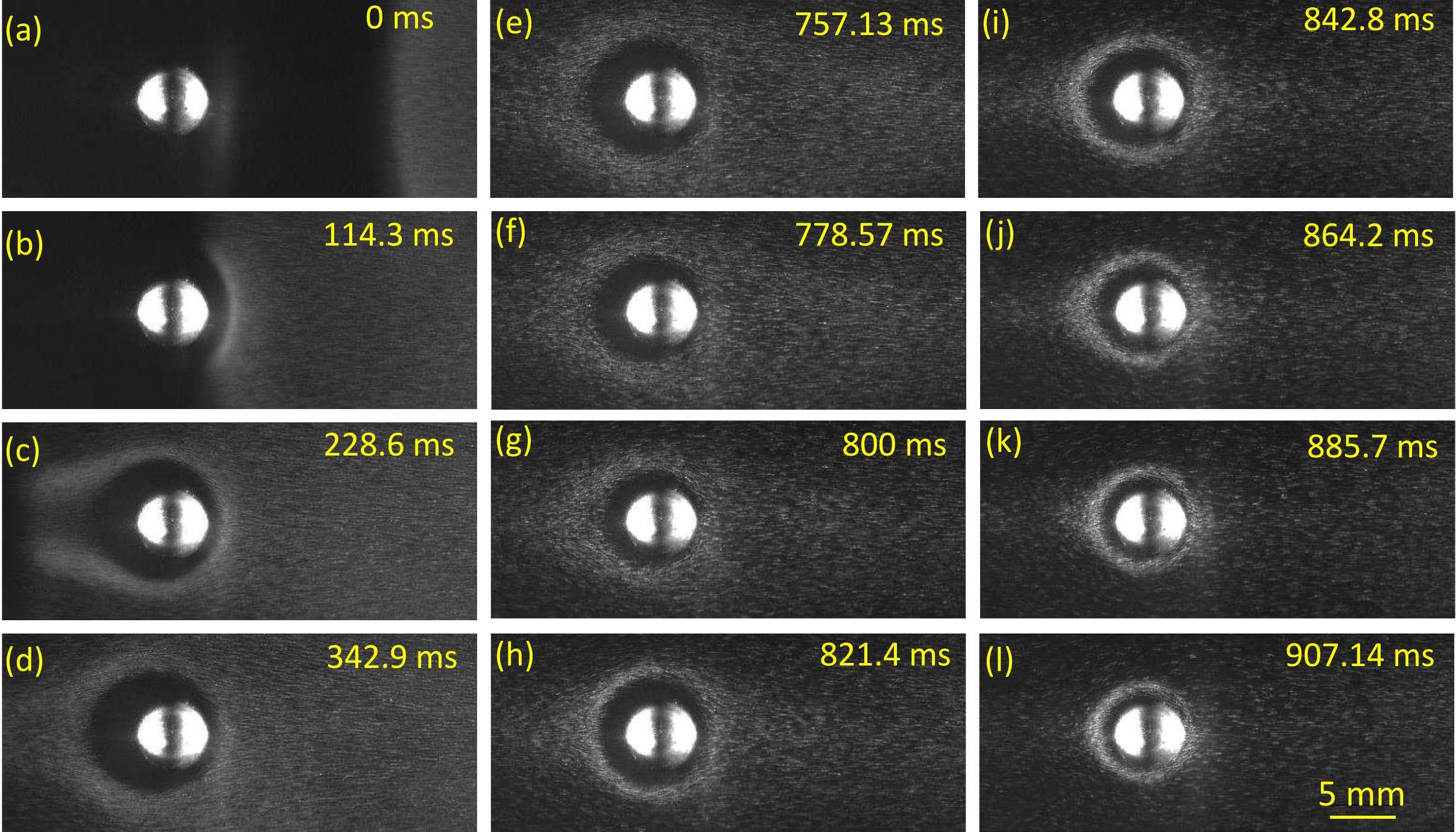}}
\caption{\label{fig:void_formation} Sequence of single frame images (in the X-Z plane) showing the dust flow around a floating spherical obstacle for a flow rate change of 11 ml$_s$/min. A small brighter patch to the immediate right of the sphere is the reflection of the laser light from the sphere. The microparticle cloud, visible as a diffuse gray region in the right corner of fig.~\ref{fig:void_formation}(a), flows from right to left with a flow velocity of 6-7 cm/sec and starts interacting with the sphere between (a) and (b).  A transient bow shaped structure appears in (b). The cloud then proceeds further in the downstream direction and forms a non-symmetric circular void at 342.9 ms (d). The cloud starts moving closer to the obstacle in between frames (e) to (g), whereas it is repelled by the sphere and forms a dust cluster in (h). The circular structure comes closer to the sphere in between frames (i)-(k) and achieves a saturation in (l).}
\end{figure*}
%%%%%%%%%%%%%%%%%%%%%%%%%%%%%%%%%%%
As discussed before, initially a stationary dust cloud is achieved by maintaining the pumping speed and gas flow rate, then allowed to flow from right to left by reducing the neutral gas flow rate suddenly from its equilibrium condition for a time duration of less than a second. When this change of background gas flow rate is very small (e.g., when the opening of mass flow controller changes from 10\% to 5\% or less) the cloud as a whole flows with very small velocity. Thus it is possible to observe its entire interaction with the obstacle, from the time of initial incidence until the cloud completely passes over the sphere. It is known from past studies that usually a dust free region forms around the potential barrier surrounding the obstacle where an equilibrium of the electric field force and the ion-drag force is established which could depend on the biasing of the obstacle \cite{Klindworth, ed}. Thus we have first started the experiment with the object at floating potential. Fig.~\ref{fig:void_formation} shows a sequence of images of the dust cloud interaction with the spherical obstacle in X-Z plane when the sphere is kept at floating potential while the flow is generated by changing the gas flow rate from its equilibrium (10\%) to 6\% (27.5 ml$_s$/min to 16.5 ml$_s$/min) in a time span of less than a second. In each image, the dust cloud flows from right to left, and the bright circle in the center of each image is the metallic ball. Upon changing the flow rate the dust particles start flowing towards the spherical obstacle with a velocity of 6 - 7 cm/sec (Fig.~\ref{fig:void_formation}(a)). It is worth mentioning that the dust cloud quickly attains an asymptotic velocity after an initial acceleration which remains uniform over a substantial region during the time of experimental observation. The final constant velocity of the dust cloud depends on the selected flow rate difference (see Ref. \cite{jaiswal4}). The moving cloud interacts with the circular obstacle and bends like a bow in the upstream region 
(Fig.~\ref{fig:void_formation}(b)) due to the repulsive force exerted by the metallic object, but continues to propagate towards the downstream direction due to neutral streaming. The deceleration of the flow due to the potential barrier caused by the obstacle results in the formation of a non symmetric circular void around the floating sphere as can be seen in Figs.~\ref{fig:void_formation}(c)-(d)). At a later stage ($\sim 757$ ms) we observe a dynamical change in the shape of the void: it gets contracted and the microparticle cloud starts getting closer to the obstacle as depicted by Figs.~\ref{fig:void_formation}(e)-(g)). However, the cloud is prevented from getting any closer than about $\sim 1.5$ mm from the metallic ball due to the strong repulsion of the charged sphere (see Fig.~\ref{fig:void_formation}(h)). This shows a threshold distance between the obstacle and the particles beyond which the particles are attracted. The strong reflection (repulsion) due to the obstacle causes an accumulation of particles in the upstream direction with a visibly enhanced density (visible as circular brighter region in Figs.~\ref{fig:void_formation}(h)-(l)). The fact that this region keeps a circular shape even in the downstream direction is a clear indication of an effective attraction towards the obstacle at intermediate distances. After its initial formation, this circular region again shrinks towards the sphere and acquires a stable spherical shape (with the obstacle displaced slightly from the center) after reaching a saturated distance that can be seen from Figs.~\ref{fig:void_formation}(i)-(l).     
     
This attraction - repulsion phenomenon is akin to past observations of boundary-free clusters by Usachev \textit{et al.} \cite{usachev_dust_cluster} and formation of a secondary void around a Langmuir probe/ metallic ball inside the primary void observed by Klindworth \textit{et al.}\cite{Klindworth} and Schwabe \textit{et al.}\cite{schwabe}. %Schwabe \textit{et al.} \cite{schwabe} termed this behavior as \lq repulsive attraction'. 
%Repulsive attraction between charged particles and a negatively biased wire has been studied by Samsonov \textit{et al.}
The attraction and repulsion between charged particles and a negatively biased wire has been studied by Samsonov \textit{et al.} \cite{samsonov3} where they have attributed the attraction of microparticles to the ion-drag force and the repulsion to the electrostatic force. Our experimental observation can also be understood in terms of the following physical picture. The metallic sphere significantly affects both the local plasma environment and the microparticle dynamics. The sphere gets charged under the influence of the plasma and repels the flowing microparticles close to it thereby forming a void in the surrounding region. However, the distant particles are attracted towards the sphere due to the ion drag force, causing an increased negative charge density around the obstacle that are repelled by the electrostatic force that dominates over the ion drag at a smaller distance. This results in the formation of a ring shaped structure around the obstacle. This circular structure stabilizes at a distance where the ion drag force equals the combined electrostatic repulsion of the charged object %, the interparticle repulsion between the dust particles 
and the effect of the gas drag. This distance of balance is different in the downstream and upstream regions.\par 
 %%%%%%%%%%%%%%%%%%%%%%%%%%%%%%%%%%%%%%%%%%%%%%%%%%%%%%%%%   
\begin{figure}[!ht]
\centering{\includegraphics[scale=0.65]{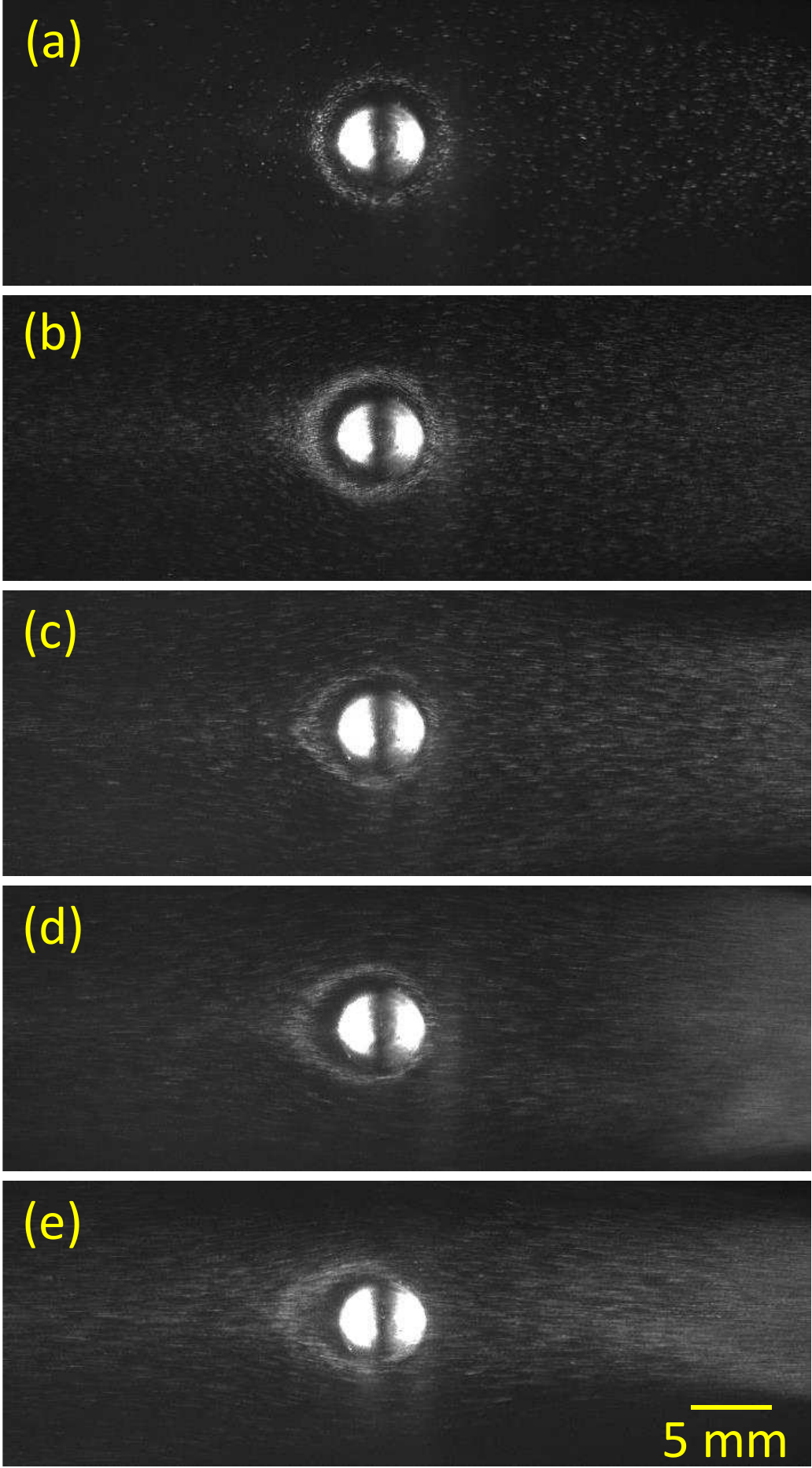}}
\caption{\label{fig:structure_vs_flow_rate} Experimental Images of microparticle flow around a floating obstacle for flow rate changes of (a) 8.25 ml$_s$/min, (b) 11 ml$_s$/min, (c) 13.75 ml$_s$/min, (d) 16.5 ml$_s$/min and (e) 19.25 ml$_s$/min. 
The circular region in the middle of the flow is the spherical metallic ball  which acts as an obstruction for the flow of dust fluid.}
\end{figure}
%%%%%%%%%%%%%%%%%%%%%%%% 
We next study the dust flow dynamics at different flow velocities (induced by varying the neutral flow) while keeping the obstacle at the floating potential.  Fig.~\ref{fig:structure_vs_flow_rate} shows single frame snapshots of the dust cloud flowing past a floating obstacle with different fluid flow velocities. The flow velocity is changed by changing the flow rate in intervals of 1\% (2.75 ml$_s$/min). As observed before, at the various flow rates, the dust cloud initially flows around the object, forms a void which then shrinks towards the sphere to form a circle around it. For further analysis, we have selected those frames  where the circular structure around the obstacle has already reached its equilibrium position.  
Fig.~\ref{fig:structure_vs_flow_rate}(a) shows a single frame depicting the structure formed at a flow rate change of 8.25 ml$_s$/min. It can be seen from the image that the dust cloud forms a complete circle with the obstacle in the middle. This means that the effect of neutral streaming is negligible compared to the ion drag force and electrostatic repulsive force of the obstacle. Consequently the dust cloud takes the shape of a stable ring like structure where the two forces counterbalance each other. Fig.~\ref{fig:structure_vs_flow_rate}(b) shows that a slight modification in the circular structure occurs at the leading edge of the void for the flow generated at a flow rate difference of 11.0 ml$_s$/min. We have found such stable structures to form around the obstacle even at a higher dust flow velocity ($v_f$).  However the structures then show an upstream/downstream asymmetry in the distribution of dust as can be seen from Fig.~\ref{fig:structure_vs_flow_rate}(c). Fig.~\ref{fig:structure_vs_flow_rate}(d) shows the structure formation at a flow rate difference of 6\% (16.5 ml$_s$/min). We can see from the image that the tail of the void gets enlarged as the asymmetry in the structure increases with the increasing value of $v_f$. For a dust cloud moving with a highly supersonic velocity at a flow rate change of 7\% (19.25 ml$_s$/min), we observe a very deformed structure that is stretched in the downstream region of the obstacle (see
Fig.\ref{fig:structure_vs_flow_rate}(e). We also find the existence of a small fluctuation in the structure in the downstream region.\par
 %%%%%%%%%%%%%%%%%%%%%%%%%%%%%%%%%%%%%%%%%%%%%%%%%%%%%%%%%   
\begin{figure}[!ht]
\centering{\includegraphics[scale=0.65]{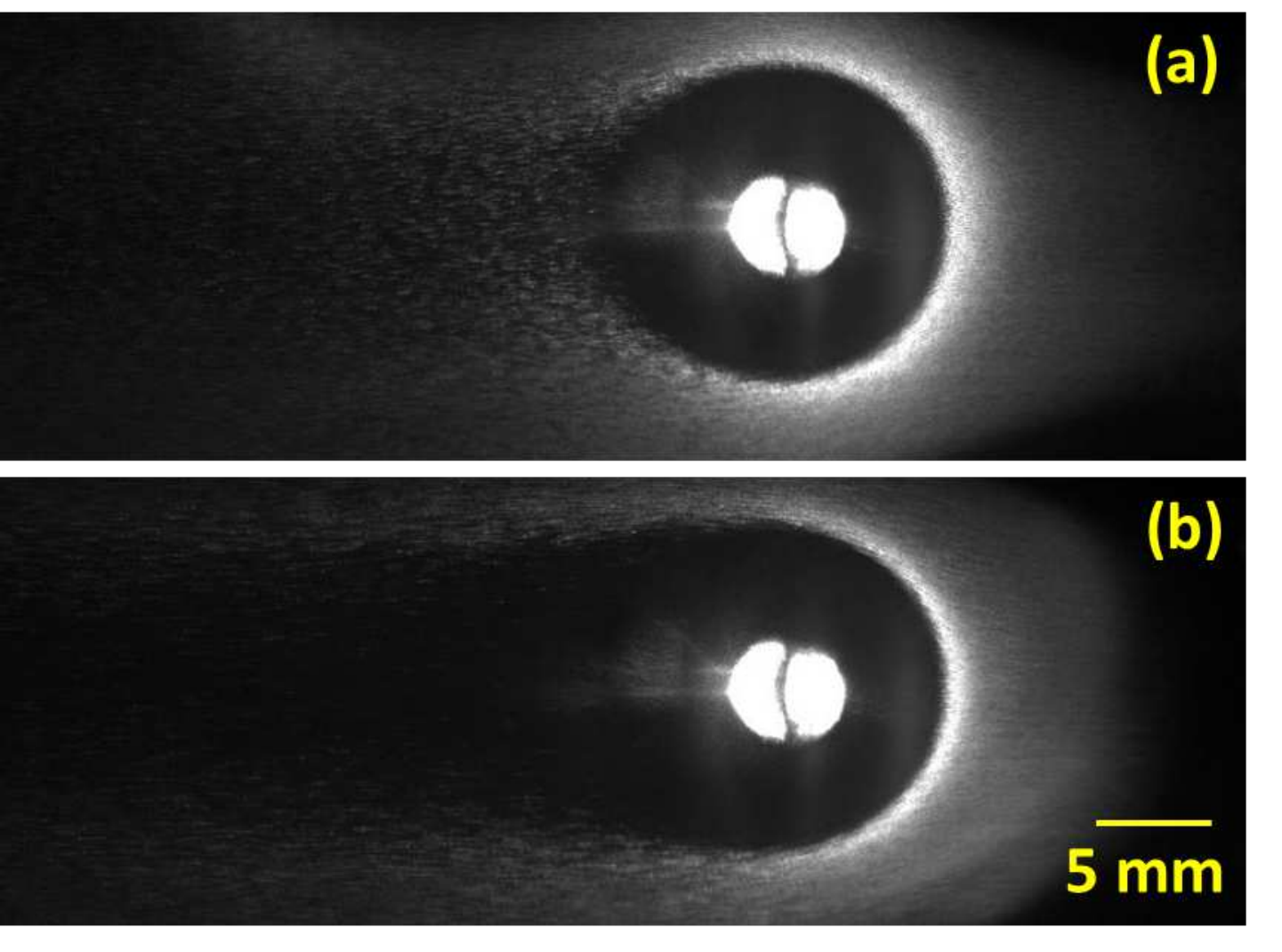}}
\caption{\label{fig:biased_vs_flow_rate1} Experimental images of microparticle flow around a biased obstacle for a flow rate change of (a) 11 ml$_s$/min, (b) 16.5 ml$_s$/min. The discharge voltage is 350 V and the dc bias applied to the obstacle is 250 V with respect to ground. The circular region in the middle of the flow is the spherical metallic ball which acts as an obstruction for the flow of dust fluid.}
\end{figure}
%%%%%%%%%%%%%%%%%%%%%%%% 
The effect of neutral streaming on the interaction of dust cloud with the spherical obstacle can be understood in a more efficient way by biasing the  obstacle at different voltages since the electrostatic repulsion due to the charged object increases with the increase of biasing voltage of the obstacle \cite{samsonov3}. Therefore, in order to check how the observed effect depends on the biasing of the obstacle, we have biased the sphere at different voltages with respect to ground. Fig.~\ref{fig:biased_vs_flow_rate1} shows the dust flow past an obstacle with different flow velocities while the object is biased at 250 V with respect to ground. The discharge voltage in this case is 350 V, and the plasma potential is measured as $\sim$ 335 V. Fig.~\ref{fig:biased_vs_flow_rate1}(a) depicts the interaction of a biased obstacle with the flowing dust cloud generated by suddenly reducing the mass flow rate from equilibrium (27.5 ml$_s$/min) to 16.5 ml$_s$/min. For this small flow rate change of 11 ml$_s$/min 
the cloud moved with relatively lower velocity (subsonic) \cite{jaiswal4} and showed similar behavior as observed for the floating obstacle. While flowing around the biased obstacle the microparticles follow the potential barrier caused by the sphere and as usual a void is formed due to the electrostatic repulsion. 
In this case, we did not observe any ring shaped structure formation around the obstacle in the downstream direction. This indicates that the attractive ion drag force is less than the combined effect of the electrostatic repulsion and neutral streaming.
 This observation is well supported by past findings of Samsonov \textit{et al.} \cite{samsonov3}. They performed an experiment to test the attraction or repulsion of distant or near particles to a wire placed in parallel to the plane of the levitating particles. The calculated total force exerted on the particles was repulsive near the wire where the electric repulsion prevailed. 
 However, it became attractive at larger distances due to an increase in the ion drag towards the wire. For a more negative biasing the total force remained repulsive, and the point where the net force was zero shifted away from the wire. Applying the same concept to our experimental result we can say that in this case the electrostatic repulsion is larger so that particle could not come close to the object and hence there was no ring formation around the obstacle. 
 For the case of a larger flow rate change of 16.5 ml$_s$/min the neutral streaming is comparatively larger than before and hence a streamline shaped void is formed during the interaction of the dust and the metallic ball as can be seen in Fig.~\ref{fig:biased_vs_flow_rate1}(b). The tail of the cloud is expanded in $+$Z - direction and the
two streams of the cloud remain separated for a longer distance in the downstream region of the obstacle.

%%%%%%%%%%%%%%%%%%%%%%%
\begin{figure*}[!ht]
 \centering{\includegraphics[width=0.8\textwidth]{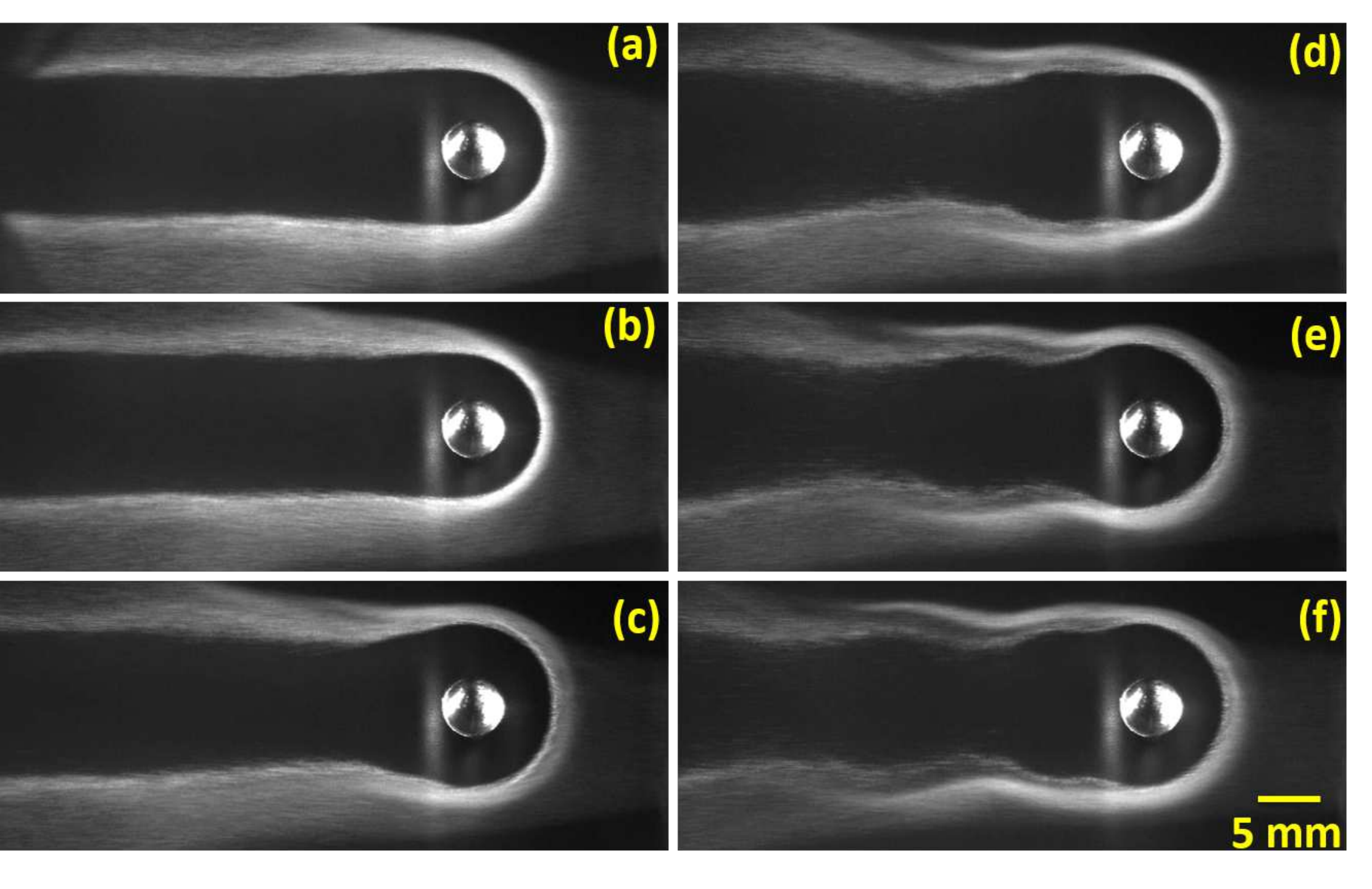}}
  \caption{\label{fig:fig5} A sequence of experimental snapshots (in the X-Z plane) at 32.8 ms intervals, showing the dust flow around a biased spherical obstacle for a flow rate of 19.25 ml$_s$/min. The metallic object is biased by applying 250 V with respect to ground. The cloud flows from the right to left with a supersonic velocity ($\approx 14~$cm/s) %of 11-12 cm/sec. 
  and interacts with the spherical obstacle in (a) and moves closer due to strong neutral streaming in (b). The cloud undergoes a strong electrostatic repulsion by the obstacle and a bow shock is formed in (c). This creates an effective perturbation in the dust cloud and a fluctuation propagates in the downstream direction in (d)-(f). The bright patch immediately to the left of the sphere in all the pictures is simply the reflected light which is subtracted as a background in further analyses.}
\end{figure*}
%%%%%%%%%%%%%%%%%%%%%%%%%%%%%%%%%%%
We then looked at experimental runs with a larger flow velocity of the dust cloud. This was done by changing the gas flow rate by 19.25 ml$_s$/min. 
In our previous experiments on flow generation techniques and measurements of flow velocity (see Ref.~\cite{jaiswal4}) we have found that for a flow rate change of more than 13.75 ml$_s$/min the dust particles (carried by the neutrals) flow towards the pump with supersonic velocity. Fig.~\ref{fig:fig5} demonstrates the interaction of the highly supersonic dust flow with a stationary obstacle biased at 250 V with respect to the ground. The potential barrier created by the biased metallic ball opposes the free motion of the particles, resulting in a compression of the cloud in the upstream side of the obstacle (see Fig.~\ref{fig:fig5}(a)). However, due to a continuous flow of neutrals which carry the dust along its way, the dust cloud gets separated while passing through the obstacle and propagates further towards the pump (from right to left). Due to the high speed, the cloud remains separated in the downstream region in a manner similar to Fig.~\ref{fig:biased_vs_flow_rate1}(b). Additionally, the compression of the dust cloud leads to an accumulation of negatively charged microparticles at the front of the obstacle and cause 
%which move close to obstacle and reflected further in upstream due to strong repulsive force by the charge sphere. This causes 
the formation of an arcuate structure in the front of the leading edge of the void as shown in Fig.~\ref{fig:fig5}(c). This also creates a strong perturbation in the dust flow which gets amplified and propagates in the downstream region of the obstacle (Fig.~\ref{fig:fig5}(d)). The downstream propagation of the fluctuation leads to undulations of the cloud edge behind the obstacle as shown in Fig.~\ref{fig:fig5}(e-f). The generation of such structures mimics the phenomenon observed during the interaction of the solar wind with the earth's magnetic field \cite{walker}.% Also, these undulations visually resemble the beginning stage of the formation of a Von k\'{a}rm\'{a}n vortex street \cite{Etienne2012} in neutral fluid flows, which is expected at Reynolds number of approximately 47 \cite{harish}.
%when the Reynolds number becomes 47.
%%%%%%%%%%%%%%%%%%%%%%%%%
%%%%%%%%%%%%%%%%%%%%%%%%%%%%%%%%%%%%%%%%%%%%%%%%%%%%%%%%%%%%   
\begin{figure}[!ht]
\centering{\includegraphics[scale=0.9]{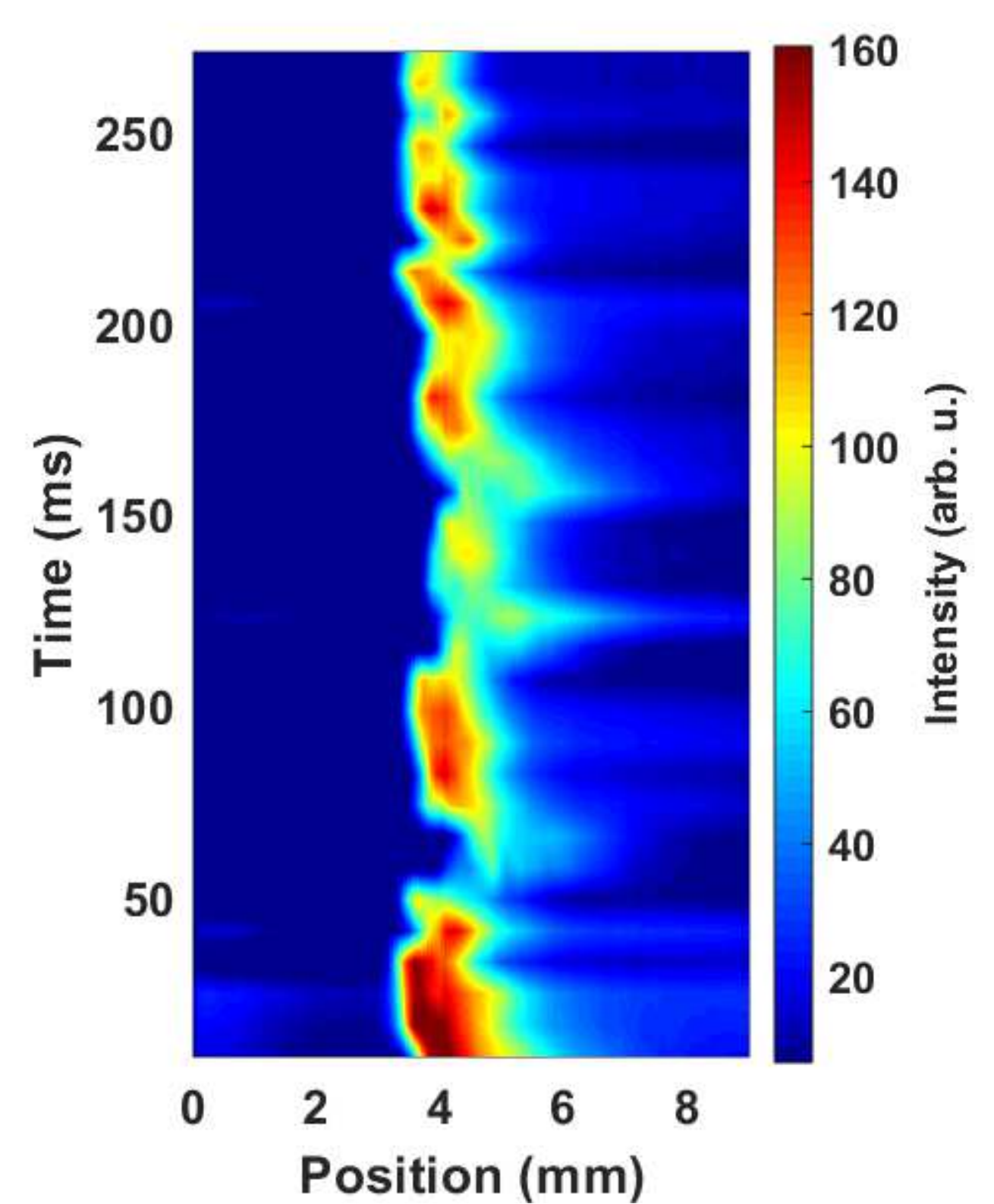}}
\caption{\label{fig:fluctuation_bow} Time evolution of the position of the leading edge of the void during interaction of supersonic dust flow with the biased obstacle. \lq\lq 0" position shows the right edge of the spherical object. Color indicates the pixel intensity in arbitrary, unnormalized units, calculated after subtracting the background intensity including any reflected light from the sphere as seen in fig.~\ref{fig:fig5}.}
\end{figure}
%%%%%%%%%%%%%%%%%%%%%%%% 
 %%%%%%%%%%%%%%%%%%%%%%%%%%%%%%%%%%%%%%%%%%%%%%%%%%%%%%%%%   
\begin{figure}[!hb]
\centering{\includegraphics[scale=0.8]{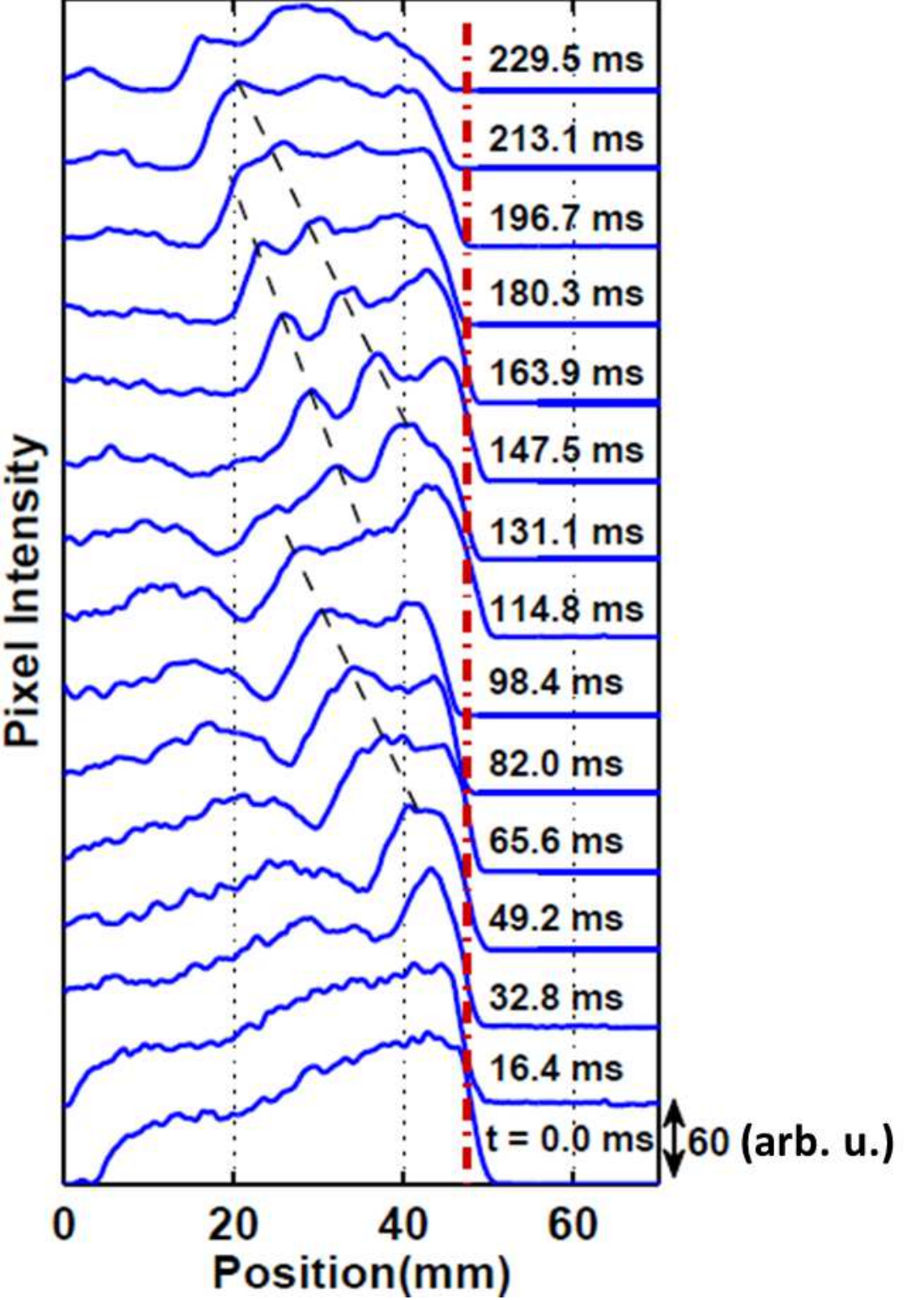}}
\caption{\label{fig:time_evolution_wave} Time evolution of the intensity of the wave crest (at intervals of 16.35 ms) generated as a consequence of fluctuations of the dust cloud in the region upstream of the biased obstacle. The dashed line represents the position of the sphere. The plots correspond to the upper front of the Fig.~\ref{fig:fig5}.}
\end{figure}
%%%%%%%%%%%%%%%%%%%%%%%% 
Fig.~\ref{fig:fluctuation_bow} shows the fluctuation of the dust cloud at the front side of the obstacle as it evolves over space and time. The \lq\lq 0" position represents the location of the sphere while the location of the void edge is close to $\approx4~$mm.  It can be seen from the figure that the position of the void edge fluctuates in space and time. A similar effect has been observed by Meyer and Merlino, who performed experiments on the interaction of a dust stream with a biased wire and observed a transient bow shock in the upstream region of the wire \cite{meyer2}. Their explanation of this phenomenon was that, as the negative charge density increased in front of the obstacle, the repulsive force was increased, pushing the dust further upstream. This causes the formation of bow shocks. %It can be seen from the plot that the fluctuation in the leading edge of the void occur at every $\sim$ 50 ms. 
%Effects that are not taken into account are the variation of the dust charge and ion flow as function of the dust %number density, among others \cite{naumkin}  
%%%%%%%%%%%%%%%%%%%%%%%%%%%%%%%%%%%%%%%%%%%%%%%%%%%%%% Previous explanation
 %As discussed above, it can be seen from the figure that due to continuous flow, dust cloud get closer to the obstacle but slowed down by the repulsive electric field. Therefore, the negative charge density increased in front of the obstacle which enhanced the repulsive force, pushing the dust further upstream at $\sim 50$ ms. This causes the formation of bow shocks which further move towards the obstacle due to strong neutral streaming and undergo the repulsion again after $\sim 100$ ms. This creates a fluctuation of leading edge of the void at every $\sim 50$ ms which get amplified and propagate as a wave structure in the downstream of the obstacle. 
The fluctuation that arises in the cloud in the upstream region subsequently propagates as a wave structure downstream of the obstacle. The wave structure is nearly symmetric around the obstacle while propagating in the downstream region. The time evolution of such a structure is depicted in Fig.~\ref{fig:time_evolution_wave}. The dashed line in the figure shows the location of the spherical obstacle. The small oscillations that develop due to the fluctuation of the leading edge of the void grows with time and splits into a number of crests which propagate in the downstream region of the obstacle.  The propagation velocity is measured as 23-26 cm/s which is approximately 10 cm/s larger than the maximum flow velocity of the dust fluid. 

  The Reynolds number $R_e = v_f L/\eta $ can be calculated for the typical void size (L) of $\sim 12~$mm (see Figs.~\ref{fig:biased_vs_flow_rate1} and \ref{fig:fig5}) and flow velocities ranging from $v_f = 4-15~$cm/sec. The kinematic viscosity $\eta $ for complex plasmas in the fluid state is $\eta = 1-10~$mm$^2$/sec. (depending on the coupling strength and density) \cite{morfill2, heideman2}. This yields a Reynolds number (for a maximum $\eta = 10~$mm$^2$/sec.) of 48 - 180 depending on variation in flow velocities, which is an order of magnitude larger than previously reported values of $R_e = 3-50$  in dusty plasmas \cite{morfill2, heideman2, schwabe2}. \par%In general, a Von k\'{a}rm\'{a}n vortex street is formed in this range of Reynolds numbers ($45\le R_e \le 10^5$) for the case of neutral fluid flow past an obstacle \cite{harish}.\par %The obtained Reynolds number in our experiment is larger than previously reported values $R_e = 3-50$ \cite{morfill2, heideman2}  
 
%As the experimental result showed that the particles coming towards the obstacle are stopped at certain distance (which varies with the biasing of the obstacle) which could be due to the balance of forces acting over on the particles, we now investigate the force balance, as in \cite{samsonov3,  usachev_dust_cluster, schwabe}, in order to calculate the distance where particle are stopped. 
%The experimental observation shows 
From the experimental results described above we see that the particles approaching the obstacle are stopped at a certain distance from the obstacle. This distance varies with the biasing voltage. This equilibrium distance is normally determined by a balance between the Coulomb repulsive force of the charged obstacle and the attractive ion drag force due to the ions which are accelerated inward by the electric field of the obstacle. However, in our experiments, since the dust cloud is made to flow by a change in the mass flow rate of neutrals, an additional force due to neutral streaming also comes into play. Thus at the void edge, the force balance can be written as $F_e = F_i+F_n$, where $F_e$ is the electrostatic repulsive force due to obstacle whereas $F_i$ is the ion drag force and $F_n$ denotes the force due to neutral streaming. In this simple model, we have neglected the effect of interparticle repulsion, which makes it possible to have several layers of microparticles arrange around the line of equilibrium between the ion drag force, electric force and neutral drag force as shown in the numerical simulation by Schwabe \textit{et al.}\cite{schwabe3}. We also neglect the effect of the dust particles on the plasma. %We neglect the interparticle repulsion and effect of the dust particles on the plasma in this simple model. 
The force on the dust due to streaming neutrals can be calculated by the Epstein formula \cite{epstein}, given as $F_{n}=-\frac{4}{3}\gamma_{Eps}{\pi}r_d^2m_nN_nv_{tn}v_f$ where, $m_n$, $N_n$, $v_{tn}$ and $v_f$ are the mass, number density, thermal and drift velocities of the neutrals respectively. $\gamma_{Eps}$ represents the Epstein drag coefficient which varies from $1$ to $1.4$ depending upon the types of reflection \cite{epstein}. The Epstein coefficient is chosen as $\gamma_{Eps} = 1.44$ \cite{jung}. The dust flow velocity ($v_f$) for the range of flow rate difference in our experiment varies as 4 - 15 cm/sec \cite{jaiswal4}. For a particle of radius 1.5 micron and Argon gas of 0.12 mbar $m_n = 6.67 \times 10^{-26}$ kg, $N_n\sim 3\times 10^{21}$ m$^{-3}$ and thermal velocity $v_{tn}$ at a temperature of 300 K, the magnitude of the force comes out to be $F_n = (0.4 - 1.6) \times 10^{-13}$ N.\par
%As the dust was slowed down by the repulsive electric field, the negative charge density increased in front of the obstacle which enhanced the repulsive force, pushing the dust further upstream. The build-up of negative charge density in front of the obstacle can be seen as an effective increase in the wire potential which provides the mechanism for the reflection of the advancing shock and the formation of the bow shock. This is analogous, in the case of the earth?s bow shock; to the compression of the magnetic field by the kinetic pressure of the solar wind.
% 
The electric force on the dust particles is given by $F_e = Q_d E$, where E is the inward electric field due to the biased sphere and $Q_d$ is the dust charge which is estimated using a rule of thumb, based on orbit motion limited (OML) theory $Q_d/e \approx 1400 T_e r_d $, as in Ref. \cite{schwabe}. The magnitude of dust charge is calculated as  $Q_d \approx 1.7 \times 10^{-15}~C$ ($Z_d \sim 1.05\times 10^4$). The OML theory can often overestimate the dust charge particularly in the case of a dense dust cloud where it is known that the charge collected on the particles decreases with increasing particle density. In order to cross check this we have followed the procedure outlined in the paper by Khrapak \textit{et al.} \cite{khrapak_charge} that allows estimating the change in the charge due to its dependence on the particle-to-plasma density ratio P ($(4 \pi \epsilon_0 r_d k_B T_e/e^2)(n_d/n_i)$). For our experimental parameters of T$_e =5$ eV, n$_d=10^{11}~m^{-3}$ and n$_i=10^{15}~m^{-3}$ we get P =0.5, which results in a dimensionless charge of $z = e^2 Z_d / 4 \pi \epsilon_0 r_d k_B T_e = 1.95$, corresponding to $Z_d \sim 1.0170\times 10^4 $. Hence, the correction of the dust charge taking into account charge reduction in the presence of dense dust clouds is less than 4\% and thus within our experimental uncertainties.
 For the calculation of the electric field we have utilized the fact that %microparticles form a cavity around the obstacle and% 
 the electric potential of the sphere inside the plasma drops exponentially with distance $x$ as it is shielded by oppositely charged particles with a shielding length $\lambda$. The electric field strength at a distance $x$, calculated by the derivative of the potential of a sphere of radius $a$ is given as $E(x) = -(\phi_0a/x^2)e^{-x/\lambda}[1+x/\lambda]$ \cite{meyer2, chen_book}, where $\phi_0$ is the potential of the obstacle with respect to the plasma potential. Since the sphere is negative with respect to plasma potential thus it is mainly shielded by ions. However the ions, accelerating towards sheath edge achieve the Bohm velocity $v_B$ which is determined by the electron temperature. Hence, ions at Bohm velocity yield $\lambda_{i} \approx \lambda_{e}$ \cite{Klindworth}.
%But in the presheath, the ions are accelerated from the ion thermal velocity and reach the Bohm velocity $v_B% at the sheath edge. Thus at the sheath edge, the ion shielding length $\lambda_{di}$ is determined by ion energy rather than ion temperature [6].Hence, ions at Bohm velocity yield $\lambda_{di} \approx \lambda_{de}$ \cite{Klindworth}
Therefore, in this case the electron Debye length ($\lambda_{e} \approx 550 \mu m$) could be taken as the effective screening length inside the void.
The plasma and floating potentials, $V_p$ and $V_f$, at the discharge parameter $V_d = 360~$V and P = 0.12 mbar as measured by emissive probe are $ V_p \simeq+(340-345)~V$ and $V_f\simeq +(325-330)~V$ respectively. Thus, from the known bias voltage of the sphere ($V_b\simeq+250~V$) and measured plasma and floating potentials, $\phi_0$ comes out to be $-(10-15$) V and $\sim -95~V$ for the case of floating and biased obstacles respectively.\par
The ion-drag force $F_i$ is estimated by using the formula given by Khrapak \textit{et al.} \cite{Khrapak_ion_drag}. The drag force for a microparticle in a collisionless Maxwellian plasma with arbitrary ion velocity is given as:
%%%%%%%%%%%%%%%%%%%%%%%%%%%
%\begin{eqnarray}
%         \label{eq:Fi}
%         \fl {\mathsub{F}{i}} = \sqrt{2 \pi} \mathsub{r}{d}^2 \mathsub{n}{i}
% \mathsub{m}{i} \mathsub{v}{Ti}^2 \left\{ \sqrt{\frac{\pi}{2}} \right.
% \text{erf} \left( \frac{u}{\sqrt{2}} \right) [1 +u + (1-u^{-2})(1+2z\tau) +
%          4 z^2 \tau^2 u^{-2} \ln \Lambda ] \nonumber\\+ \left. u^{-1} [1 + 2 z
% \tau + u^2 - 4 z^2 \tau^2 \ln \Lambda]  \exp \left( - \frac{u^2}{2}
% \right) \right\},
% \end{eqnarray}
%%%%%%%%%%%%%%%%%%%%%%%%%%%%
\begin{eqnarray}
F_i = \sqrt{2\pi}r_d^2n_im_iv_{ti}^2 \left\{ \sqrt{\frac{\pi}{2}} \right.
\text{erf} \left( \frac{u_f}{\sqrt2} \right) 
[ 1+ u_f + (1-u_f^{-2})(1+2z\tau) + 4z^2 \tau^2 u_f^{-2} \text{ln} \Lambda ]\nonumber\\
 + \left. u_f^{-1} \bigl[ 1+2z\tau+u_f^{2}-4z^2\tau^2\text{ln}\Lambda \bigr] \text{exp}\left(-\frac{u_f^2}{2}\right) \right\}
\label{eqn:ion_drag}
\end{eqnarray}
%%%%%%%%%%%%%%%%%%%%%%%%%%%%%
where $r_d$ is the dust radius, $n_i$ and $m_i$ are the ion number density and mass, $v_{ti} = \sqrt{8k_BT_i/m_i\pi}$ is the ion thermal velocity and $u_f$ is the ion speed normalized with the ion thermal velocity. $z = Q_d|e|/4\pi\epsilon_0r_dk_BT_e$ is the normalized charge whereas $\tau = T_e/T_i$. ln$\Lambda$ is the Coulomb logarithm which is given as:
\begin{equation}
\text{ln}\Lambda = \text{ln}\left[\frac{\beta+1}{\beta+(r_d/\lambda_{ef})}\right]\nonumber
\end{equation}
where $\beta$ is defined as $\beta = Q_d|e|/(4\pi\epsilon_0k_BT_i(1+u_f^2)\lambda_{ef})$ and $\lambda_{ef}(u_f) = 1/\sqrt{\lambda_i^{-2}(1+u_f^2)^{-1}+\lambda_e^{-2}}$ is the effective screening length. The ion flow velocity is calculated by using the formula $v_i = \mu_i E$. E is inward electric field due to charge obstacle and $\mu_i$ is the ion mobility, estimated by using the expression $\mu_i(E) = \mu_0/p\sqrt{(1+\alpha E/p)}$ with $\mu_0 = 19.5~$m$^2$ Pa V$^{-1}$ s$^{-1}$, $\alpha = 0.035~$m Pa V$^{-1}$ for argon gas of pressure p (in Pascal) \cite{schwabe}. 
For the experimental parameters mentioned above the obtained ion drag force becomes almost equal to the electrostatic force (and $F_i+F_n = F_e$) at a distance of $\sim1.57$ mm in the case of a floating obstacle. The electric field strength at this position is calculated as E = 2284 V/m which results in an ion drag force of $F_i = 3.78\times 10^{-12}$ N and electrostatic force $F_e = 3.8 \times 10^{-12}$ N. The estimated distance is quite close to the experimentally observed distance of the void edge from the obstacle. Further, the observed asymmetry in the leading and trailing edge of the void may be due to contributions from the neutral streaming effect. Although this effect is much smaller than the other two forces it may be sufficient to cause this small asymmetry  by adding on to the ion drag force in the upstream region.   At larger distances (larger than the equilibrium position) $F_i/Fe > 1$, corresponding to the attraction of particles towards the sphere, while closer to the obstacle, the electrostatic interaction is larger ($F_i/Fe < 1$), so that the total force is repulsive. In the case of the biased obstacle the equilibrium distance is estimated to be $\sim 2.5$ mm, which is comparable with the measured distance of $\sim 2.7 - 3.2~$mm separating the leading void edge from the obstacle. For a high flow rate change of $19.25~$ml$_s$/min the dust flow velocity is high enough to cross the equilibrium distance and to experience a strong electrostatic repulsion. As a result the void around the obstacle now blows apart and leads to the formation of a bow shock in the upstream region while a small fluctuation grows and moves in downstream region of the obstacle. 
Our calculated values, presented above, provide a good approximate estimate of the observed experimental data. Further refinements in the theoretical estimates can be effected by taking into account  the  variation of the screening length inside the void as a function of position to calculate the electric field more accurately. Also, the charge of the microparticles has been approximated. A more precise determination as a function of the dust number density, would also improve the accuracy of the force balance. We have also not taken into account the influence of the dust and the streaming background gas on the ion drag. In general, a self-consistent simulation of the plasma, microparticles and sphere, including plasma fluxes and charging processes, would be a worthwhile theoretical simulation project to carry out an in depth exploration of the underlying physics of the phenomena reported in our experiments. 
%%%%%%%%%%%%%%%%%%%%%%%%%%%%%%%%%5
%%%%%%%%%%%%%%%%%%%%
\section{Conclusion}
\label{sec:conclusion}
In conclusion, an experimental study of the flow of a dusty plasma past a metal spherical obstacle is presented. Observations are made for various flow velocities with the obstacle at a floating potential as well as when it is negatively biased with respect to the plasma.  A void is formed when the flowing dust particles are decelerated in the vicinity of the obstacle due to its repulsive force. When the obstacle is kept at a floating potential, the distant dust particles are attracted towards the obstacle due to the ion drag force, whereas they are repelled when they get very close to the negatively charged sphere (at a distance $\lesssim 1.5~$mm). This interplay of forces results in the formation of a ring shaped structure around the obstacle whose shape varies with the change of flow velocities. In the case of a negatively biased obstacle, a streamline shaped void is formed when the dusty plasma flow velocity is relatively low compared to the dust acoustic speed. For a supersonic flow of the dust cloud a bow shock is excited in the upstream direction of the obstacle while small density perturbations that grow and transform into wave structures are seen to propagate in the downstream region. The fact that the undulations of the cloud edge behind the obstacle only appear when the obstacle is sufficiently negatively biased and the flow is supersonic clearly indicates  the existence of thresholds for the applied obstacle bias and the dust flow velocity for the generation of such structures, and also demonstrates that the undulations are not caused by the neutral gas since it is unaffected by the obstacle bias. A force balance relation between the electrostatic force, the ion drag force and the neutral streaming effect is used to calculate the distance of separation between the obstacle and the edge of the void. Our theoretical estimates agree quite well with the experimental observations. The dynamical behavior of the dust flow as a function of the flow velocity and the obstacle bias, as seen in our experiments, could provide useful insights for understanding related phenomena wherever flowing charged particles interact with a stationary body. Such situations abound in many laboratory and industrial applications of plasmas as well as on a larger scale in extraterrestrial and astrophysical scenarios. \\
 %%%%%%%%%%%%%%%%%%%%%%%%%%%%%%%%%%%%%%%%%%%%%%%%%%%%%%%%%
\section{acknowledgement}
Authors S. Jaiswal and M. Schwabe would like to thank S. K. Zhdanov for helpful discussions, and J. Meyer for carefully reading the manuscript and for helpful comments. S. Jaiswal acknowledges the support of DLR-DAAD Research Fellowships.
\section*{References}
%###################################################################################

\end{document}